\input psfig.sty
\def\ptitle{Spectral bounds for the cutoff Coulomb potential}
\nopagenumbers
\magnification=\magstep1
\hsize 6.0 true in 
\hoffset 0.25 true in 
\emergencystretch=0.6 in                 
\vfuzz 0.4 in                            
\hfuzz  0.4 in                           
\vglue 0.1true in
\mathsurround=2pt                        
\topskip=24pt                            
\def\nl{\noindent}                       
\def\np{\hfil\vfil\break}                
\def\title#1{\bigskip\noindent\bf #1 ~ \tr\smallskip} 
\font\tr=cmr10                          
\font\bf=cmbx10                         
\font\sl=cmsl10                         
\font\it=cmti10                         
\font\trbig=cmbx10 scaled 1500          
\font\tiny=cmr8                         
\def\ng{>\kern -9pt|\kern 9pt}          
\def\hi#1#2{$#1$\kern -2pt-#2}          
\def\hy#1#2{#1-\kern -2pt$#2$}          

\def\sgn{\rm sgn}
\def\half{{1 \over 2}}

\def\htab#1#2{{\hskip #1 in #2}}

\output={\shipout\vbox{\makeheadline
                                      \ifnum\the\pageno>1 {\hrule}  \fi 
                                      {\pagebody}   
                                      \makefootline}
                   \advancepageno}

\headline{\noindent {\ifnum\the\pageno>1 
                                   {\tiny \ptitle\hfil page~\the\pageno}\fi}}
\footline{}
\newcount\zz  \zz=0  
\newcount\q   
\newcount\qq    \qq=0  

\def\pref #1#2#3#4#5{\frenchspacing \global \advance \q by 1     
    \edef#1{\the\q}
       {\ifnum \zz=1 { %
         \item{[\the\q]} 
         {#2} {\bf #3},{ #4.}{~#5}\medskip} \fi}}

\def\bref #1#2#3#4#5{\frenchspacing \global \advance \q by 1     
    \edef#1{\the\q}
    {\ifnum \zz=1 { %
       \item{[\the\q]} 
       {#2}, {\it #3} {(#4).}{~#5}\medskip} \fi}}

\def\gref #1#2{\frenchspacing \global \advance \q by 1  
    \edef#1{\the\q}
    {\ifnum \zz=1 { %
       \item{[\the\q]} 
       {#2}\medskip} \fi}}

 \def\sref #1{~[#1]}

\def\references#1{\zz=#1
   \parskip=2pt plus 1pt   
   {\ifnum \zz=1 {\noindent \bf References \medskip} \fi} \q=\qq

\pref{\meh}{C.H. Mehta and S.H. Patil, Phys. Rev. A}{17}{43 (1978)}{}
\pref{\ray}{P. P. Ray and K. Mahata, J. Phys. A}{22}{3161 (1989)}{} 
\bref{\stegan}{M. Abramowitz and I. Stegun}{Handbook of Mathematical Functions}{Dover, New York, 1968}{}
\bref{\reed}{M. Reed and B. Simon}{Methods of Modern Mathematical Physics IV: Analysis of Operators}{Academic, New York, 1978}{The min-max principle for the discrete spectrum is discussed on p75}

\bref{\reeda}{M. Reed and B. Simon}{Methods of Modern Mathematical Physics IV: Analysis of Operators}{Academic, New York, 1978}{Theorem XII.6,a on p87}

\pref{\rhalla}{R. L. Hall, J. Math. Phys.} {24} {324 (1983)}{}

\pref{\znojila}{M. Znojil, Phys. Lett. A} {94} {120 (1983)}{}

\pref{\rhallb}{R.L Hall, J. Math. Phys. }{25} {2708 (1984)}{}

\pref{\rhallc}{R.L. Hall, J. Math. Phys.} {34} {2779 (1993)}{}
    
\pref{\rhalle}{R.L. Hall, Phys. Rev. A}{50}{2876 (1994)}{} 

\pref{\znojilb}{M. Znojil, J. Phys. A} {29} {6443 (1996)}{}

}

 \references{0}    
\vskip 1true in
\htab{3.5}{CUQM-90}

\htab{3.5}{math-ph/0201021}

\htab{3.5}{January 2002}

\tr 
\vskip 1.0true in
\centerline{\trbig Spectral bounds for the cutoff Coulomb potential  }
\vskip 0.5true in
\baselineskip 12 true pt 
\centerline{\bf Richard L. Hall and Qutaibeh D. Katatbeh}\medskip
\centerline{\sl Department of Mathematics and Statistics,}
\centerline{\sl Concordia University,}
\centerline{\sl 1455 de Maisonneuve Boulevard West,}
\centerline{\sl Montr\'eal, Qu\'ebec, Canada H3G 1M8.}
\vskip 0.2 true in
\centerline{email:\sl~~rhall@mathstat.concordia.ca}
\bigskip\bigskip

\baselineskip = 18true pt  
\centerline{\bf Abstract}\medskip
The method of potential envelopes is used to analyse the bound-state spectrum of the Schr\"odinger Hamiltonian $H=-\Delta-v/(r+b)$, where $v$ and $b$ are positive. We established simple formulas yielding upper and lower energy bounds for all the energy eigenvalues.

\medskip\noindent PACS~~31.15.Gy, 31.15.Pf, 03.65.Ge.

\np
  \title{1.~~Introduction}

The cutoff Coulomb potential $f(r)$ given by
$$ f(r)=-v/(r+b)\eqno{(1.1)}$$
 is an approximation to the potential due to a smeared charge distribution, rather than a point charge, and is appropriate for describing mesonic atoms\sref{\meh}. 
Many authors have studied the eigenvalues $E_{n\ell},  n=1,2,3,..., \ell=0,1,2,...$ generated by the cutoff Coulomb potential and have tried to estimate them. For example Ray and Mahata\sref{\ray} applied the method of \hy{large}{N} expansion to approximate the bound states energies from $n=1$ to $n=4$. Mehta and Patil\sref{\meh} rigorously analysed the S-wave bound-state eigenvalues of this potential as a function of $b$. 

 In this paper we offer an elementary proof that the cutoff Coulomb potential has infinitely many discrete negative eigenvalues $E_{n\ell},  n=1,2,3,..., \ell=0,1,2,...$ by using the comparison methods. We then use the comparison theorem and the envelope method\sref{\reed-\rhallc} to obtain simple upper- and lower-bound formulas for all the eigenvalues.
  \title{2.~~The discrete spectrum : Scaling}
The Hamiltonian for the problem is given by,
$$H=-\Delta-v/(r+b),~v,~b>0. \eqno{(2.1)}$$

 \noindent A concern might be that, for sufficiently small coupling $v$, the potential, like a square well, might not have any discrete spectrum.  However, the Coulomb tail averts this problem. It has been proved \sref{\reeda} by general methods that for any potential, like
 $-v/(r+b),$  which is negative and decays at infinity slower than $1/r^{2-\epsilon}$, the corresponding Hamiltonian  operator has infinitly many negative eigenvalues. The specific result for our problem may also be obtained by an elementary application of the comparison theorem, as we now show by the following argument. We note that the potential can be written 

$$f(r)={-v\over r}+{vb\over r^2}-{vb^2\over{r^2(r+b)}}.\eqno{(2.2)}$$
It therefore follows that
$$-{v\over r} < f(r) < {-v\over r}+{vb\over r^2},$$and consequently
$$V_{l}=-{v\over r}+{{\ell(\ell+1)}\over{r^2}}<f(r)+{{\ell(\ell+1)}\over{r^2}}
< {-v\over r }+{\lambda(\lambda+1)\over r^2} = V_{u}, \eqno{(2.3)}$$where
 $$\lambda=((\ell+{1\over 2})^2+vb)^{1\over 2}-{1\over 2}.\eqno{(2.4)}$$ Hence,
 we see that the effective potential associated with $f(r)$ is bounded above and below by Hydrogenic effective potentials with discrete negative eigenvalues. 
This implies that the potential $V$ has infinitely many negative discrete eigenvalues $E_{n\ell}$ bounded by 
$${-v^2\over{4(n+\ell)^2}} \le E_{n\ell}\le {-v^2\over{4(n+\lambda)^2}}.\eqno{(2.5)}$$    
These bounds are asymptotically close for large $n.$
 Another upper bound is provided by the linear potential since  
$f(r)<-{v\over b} +{v \over b ^2} r.$ Hence,
$$E_{n\ell}<-{v\over b}+({v\over b ^2})^{2\over 3}{\cal E}_{n\ell}(1),\eqno{(2.6)} $$ 

\noindent where ${\cal E}_{n\ell}(1)$ are the eigenvalues of the Hamiltonian $-\Delta +r $ for linear potential.
  
 For the S-states the radial equation may be transformed into Whittaker's equation which has known exact solutions \sref{\stegan}.  The general solution is written\sref{\stegan} in terms of the confluent hypergeometric functions $M[x,y,z]$ and $U[x,y,z]$ where,
$$U(x,y,z)={1\over \Gamma (x)}\int\limits_0^\infty {e^{-zt}t^{x-1}(1+t)^{y-x-1}dt}=z^{-x}{}_2F_0[x,1+x-y;;-1/z] \eqno{(2.7)} $$
\noindent and $M[x,y,z]={}_1F_1[x;y;z].$
Mehta and Patil \sref\meh ~ used the bounded property of the radial wave function and the boundary conditions to demonstrate that the eigenvalues are determined by the equation  
$$U[1-v/(2\sqrt{-E}),2,2b\sqrt{-E}]=0. \eqno{(2.8)}$$
As an alternative, we shall apply the envelope method to approximate all the eigenvalues. We first reduce the complexity of the problem by the use of scaling arguments.
 If we denote the eigenvalues of $H=-\omega \Delta -v/(r+b)$ by ${\cal E}(\omega,v,b)$, and consider a scale change of the form $s=r/{\sigma}$, and choose the scale ${\sigma}={\omega}/v,$ then it is easy to show that, \nl
$$ {\cal E}(\omega,v,b)={{v^2}\over \omega}{\cal E}(1,1,{v b \over \omega }).\eqno{(2.9)} $$
 
 \noindent Hence, the full problem is now reduced essentially to the simpler 1-parameter problem
$$ H=-\Delta - 1 /(r+b), \quad{\cal E}={\cal E}(b), b> 0. \eqno{(2.10)}$$

  \title{3.~~Energy bounds by the  Envelope Method}

 The Comparison Theorem of quantum mechanics tells us that an ordering between potentials implies a corresponding ordering of the eigenvalues. The `envelope method' is based on this result and provides us with  simple formulas for lower and upper bounds\sref{\rhalla-\rhallc}. We need a solvable model which we can use as an envelope basis. The natural bases to use in the present context are the hydrogenic and linear potentials
$$  h(r)=\sgn(q)r^q,\quad{\rm where}\quad q=-1,1.  \eqno{(3.1)} $$
\nl ~~The spectrum generated by the potential $h(r)$ may be represented exactly
by the semi-classical expression 
$${\cal E}_{n\ell}(v) = \min_{s > 0}\{s + v\bar{h}_{n\ell}(s)\},\eqno{(3.2)}$$
where the `kinetic potential' $\bar{h}_{n\ell}(s)$ associated with the  power-law potentials (3.1) are given by \sref\rhallc
$$\bar{h}(s)=(2/q)|q{\cal E}_{n\ell}^{(q)}/{(2+q)}|^{(q+2)/2}s^{-q/2},\eqno{(3.3)} $$
and ${\cal E}_{n\ell}^{(q)}$ is the exact eigenvalue of $-\Delta+\sgn (q)r^q, $ that is to say, corresponding to the pure-power potential with coupling 1.
If we now consider a potential, such as $f(r)$, which is a smooth transformation $f(r) = g(h(r))$
of $h(r),$ then it follows that a useful approximation for the corresponding kinetic potential 
$\bar{f}_{n\ell}(s)$ is given by
$$ \bar{f}_{n\ell}(s)\approx g(\bar{h}_{n\ell}(s)) \eqno{(3.4)}$$ If g is convex in (3.4),
 we get \sref{\rhalla-\rhallc} lower bounds ($\simeq = \ge  $) for all n and $\ell,$ and if g is concave we get upper bounds ($\simeq = \le $) for all n and $\ell$. 
 
For the cutoff Coulomb potential, if we use the potential $ h=-1/r$ as an envelope basis, then $g$ is convex.
 An elementary calculation shows in this case that
$$g''(h)={2v b \over({b\over r}+1)^3}>0 .\eqno{(3.5)}  $$  
And if we use the potential $h=r$ as an envelope basis, then $g$ is  concave, in fact
$$g''(h)={-2v \over(b+r)^3} <0.\eqno{(3.6)}  $$ 
 Thus in this application of the method we obtain upper energy bounds if we use $h=-1/r$ and lower energy bounds if we use $h=r.$ The following remarks explain briefly how these results are obtained.    

We suppose for definiteness that the transformation $g(h)$ is smooth and {\it convex} $\it{i.e}$ $g''>0$, then each tangent ${f^{(t)}}(r)$ to $g$ is an affine transformation of $h$ satisfying
  
$$ {f^{(t)}}(r)=a(t)+b(t)h(r)<f(r),\eqno{(3.7)}$$
where the variables  $a(t)$ and $b(t)$ are given by solving the contact equations
 
$$f(t)= a(t)+b(t)h(t)\eqno{(3.8)}$$
$${\rm and}\quad f'(t)=b(t)h'(t), \eqno{(3.9)}$$
which mean that the `tangential potential', $f^{(t)}(r)$, and its derivative agree with 
$f(r)$ at the point of contact, $r=t.$ 
The potential inequality (3.7) implies the spectral inequality 
$$E_{n\ell}(v)\ge va(t)+{\cal E}_{n\ell}(vb(t)). \eqno{(3.10)}  $$
The optimal lower bound thus obtained may then eventually\sref{\rhallc} be re-written 
$$E_{n\ell}\ge \min_{s>0}\left\{s+g(\bar{h}_{n\ell}(s))\right\}.\eqno{(3.11)} $$
In the complementary case where g is concave, the inequalities are reversed and one obtains upper bounds.

 For the power-law potentials $h(r)=\sgn(q)r^q$ we can simplify (3.11) by changing the minimization variable $s$ to $r$ defined in each case by the equation $\bar{h}_{n\ell}(s)=h(r)$ so that $g(h(r))=f(r)={-v\over (r+b)}$ and the minimization (3.2), which yields eigenvalue approximations
 for the Hamiltonian $H = -\omega\Delta + f(r),$ where $\omega > 0,$  can be expressed in the form
 $$E_{n\ell}\approx\min_{r>0}\left\{\omega{{P_{n\ell}^2(q)\over r^2}-{v\over (r+b)} }\right\}.\eqno{(3.12)} $$
\noindent We obtain a lower bound for $P_{n\ell} = P_{n\ell}(-1)=(n+\ell),$
an upper bound for $P_{n\ell} = P_{n\ell}(1),$ and a good approximation 
with the mean value $P_{n\ell} = P_{n\ell}^{M} = \half(P_{n\ell}(-1)+ P_{n\ell}(1)).$
These \hi{P}{numbers} are provided in Table (1). 

A natural question to ask is whether there exists a set of numbers $\{P_{n\ell}\}$ such that $E_{n\ell}=\min_{r>0}\left\{ {{P_{n\ell}^2\over r^2}+f(r)}\right\}$ {\it exactly.} We can see that the answer is ``no'' by an argument based on the 
 `concentration lemma'\sref\rhalle, ~which provides us with the relation between the concentration of the ground-state wave function and the size of the coupling constant $v$. More precisely, the wave function becomes more concentrated near the origin as $v$ increases. Since for large values of the coupling $v$ the ``linear'' upper bound (3.12) is very accurate (concentration near $r=0$), if there were one ``exact'' $P_{10}$, it would have to be  the linear potential value $P_{10}=P_{10}(1).$ But our upper bound is clearly above $E_{n\ell}$ for small values of $v$. Hence there are no such ``exact'' $P_{n\ell}.$

  \title{4.~~Results and conclusion}

We have derived a simple formula (3.12) for lower and upper bounds to the eigenvalues for the cutoff Coulomb potential.  In Fig.(1) we plot the eigenvalue when $(n,\ell)=(1,1)$ as a function of $b$ for the case $v = 1,$  accurate numerical values (dashed line), and our approximation with the average 
value $P_{n\ell} = \half (P_{n\ell}(-1)+P_{n\ell}(1))$ as stars. 

If we fix $b$ and consider the Hamiltonian $H=-\Delta +vf(r),$ with eigenvalues ${\cal E}(v)$, then from (3.12) we  obtain the following explicit parametric equations for the corresponding approximate energy curve $\{v, {\cal E}(v)\},$ namely
 
$${\eqalign{ v&={2(P_{n\ell})^2\over {r^3f'(r)}}\cr
{\cal E}(v)&={(P_{n\ell})^2\over r^2}+{2(P_{n\ell})^{2}f(r)\over{r^3f'(r)}}.\cr}}\eqno{(4.1)}$$

\noindent These parametric equations yield upper bounds when $P_{n\ell}=P_{n\ell}(1)$ lower bounds when $P_{n\ell}=(n+\ell),$ and a good approximation when we use the arithmetic average of $P_{n\ell}(-1)$ and $P_{n\ell}(1)$. It is interesting, perhaps, that all these curves are scaled
versions of any one of them; it is unknown if such a symmetry is true for the
corresponding exact curves.  In Fig.(2) we exhibit the graphs of the function ${\cal E}(v)$ for $b=1$ along with accurate numerical data shown as a dashed curve.  The main 
point of this work is to show that by elementary geometric reasoning one can obtain
 simple semi-classical approximations for the eigenvalues.  These results are 
complementary to purely numerical solutions and have the advantage that they
 are expressed simply and analytically and therefore allow one to explore the
 parameter space of the problem.\smallskip  
 
   \title{Acknowledgments}
Partial financial support of this work under Grant No.GP3438 from the Natural Sciences and Engineering Research Council of Canada is gratefully acknowledged. We are indebted to Professor M. Znojil for his helpful comments.\bigskip   
\np
\references{1}
\np
\noindent {\bf Table 1}~~The `input' \hi{P}{values} $P_{n\ell}^{L},$ $P_{n\ell}^{U},$ 
and the mean $P_{n\ell}^{M} = \half(P_{n\ell}(-1) + P_{n\ell}(1))$ used in the 
general formula (3.12).  
\baselineskip=16 true pt 
\def\vr{\vrule height 12 true pt depth 6 true pt}
\def\vra{\vr\hfill} \def\vrb{\hfill &\vra} \def\vrc{\hfill & \vr\cr\hrule}
\def\vrq{\vr\quad} 

$$\vbox{\offinterlineskip
 \hrule
\settabs
\+ \vrq \kern 0.4true in &\vrq \kern 0.4true in &\vrq \kern 0.9true in &\vrq \kern 0.9true in &\vrq \kern 0.9true in &\vr\cr\hrule
\+ \vra $n$ \vrb $\ell$\vrb $P_{n\ell}^{L}= n+\ell$\vrb $P_{n\ell}^{M}$\vrb $P_{n\ell}^{U}$ \vrc
\+ \vra 1\vrb 0\vrb 1\vrb 1.18804\vrb 1.37608\vrc
\+ \vra 2\vrb 0\vrb 2\vrb 2.59065\vrb 3.18131\vrc
\+ \vra 3\vrb 0\vrb 3\vrb 3.99627\vrb 4.99255\vrc
\+ \vra 4\vrb 0\vrb 4\vrb 5.40257\vrb 6.80514\vrc
\+ \vra 5\vrb 0\vrb 5\vrb 6.80911\vrb 8.61823\vrc
\+ \vra 1\vrb 1\vrb 2\vrb 2.18596\vrb 2.37192\vrc
\+ \vra 2\vrb 1\vrb 3\vrb 3.57750\vrb 4.15501\vrc
\+ \vra 3\vrb 1\vrb 4\vrb 4.97650\vrb 5.95300\vrc
\+ \vra 4\vrb 1\vrb 5\vrb 6.37850\vrb 7.75701\vrc
\+ \vra 5\vrb 1\vrb 6\vrb 7.78204\vrb 9.56408\vrc
\+ \vra 1\vrb 2\vrb 3\vrb 3.18509\vrb 3.37018\vrc
\+ \vra 2\vrb 2\vrb 4\vrb 4.57067\vrb 5.14135\vrc
\+ \vra 3\vrb 2\vrb 5\vrb 5.96455\vrb 6.92911\vrc
\+ \vra 4\vrb 2\vrb 6\vrb 7.36257\vrb 8.72515\vrc
\+ \vra 5\vrb 2\vrb 7\vrb 8.76298\vrb 10.52596\vrc
\+ \vra 1\vrb 3\vrb 4\vrb 4.18461\vrb 4.36923\vrc
\+ \vra 2\vrb 3\vrb 5\vrb 5.56649\vrb 6.13298\vrc
\+ \vra 3\vrb 3\vrb 6\vrb 6.95652\vrb 7.91304\vrc
\+ \vra 4\vrb 3\vrb 7\vrb 8.35118\vrb 9.70236\vrc
\+ \vra 5\vrb 3\vrb 8\vrb 9.74874\vrb 11.49748\vrc
\+ \vra 1\vrb 4\vrb 5\vrb 5.18431\vrb 5.36863\vrc
\+ \vra 2\vrb 4\vrb 6\vrb 6.56366\vrb 7.12732\vrc
\+ \vra 3\vrb 4\vrb 7\vrb 7.95074\vrb 8.90148\vrc
\+ \vra 4\vrb 4\vrb 8\vrb 9.34260\vrb 10.68521\vrc
\+ \vra 5\vrb 4\vrb 9\vrb 10.73766\vrb 12.47532\vrc
}$$

\np
\vskip 0.4in
~~

\hbox{\vbox{\psfig{figure=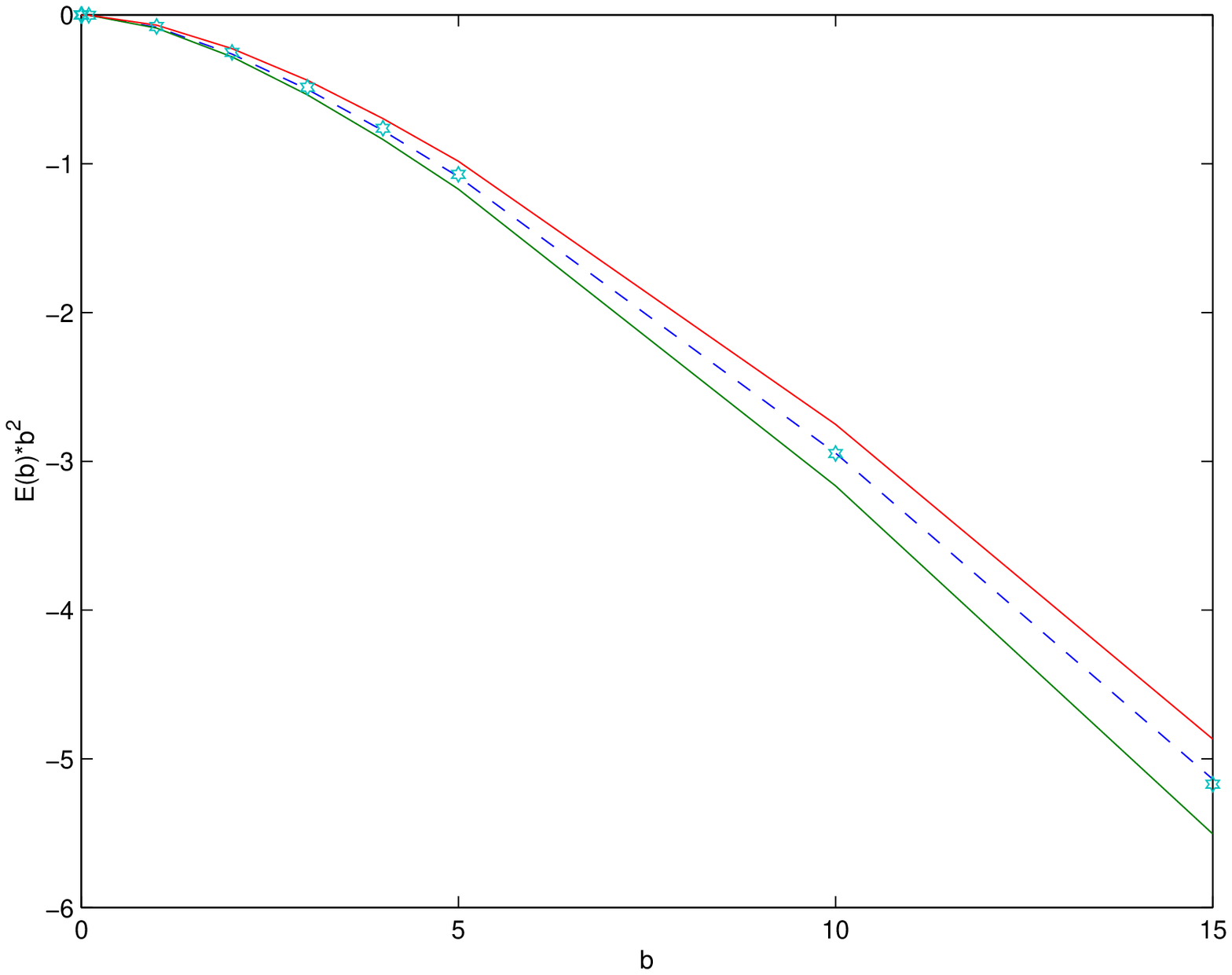,height=4in,width=3.5in,silent=}}}

\nl{\bf Figure 1.}~~The eigenvalues ${\cal E}(b)$ of the Hamiltonian $H=-\half\Delta-1/(r+b)$ for $n=\ell=1$ (in atomic units $\hbar = m = 1$). The continuous curves show the bounds given by formula (3.12), the dashed curve  represents accurate numerical data, and the stars are the `mean approximation'
 $P_{n\ell} = \half(P_{n\ell}(-1)+P_{n\ell}(1)).$\medskip

\np
\vskip 0.4in 
~~

\hbox{\vbox{\psfig{figure=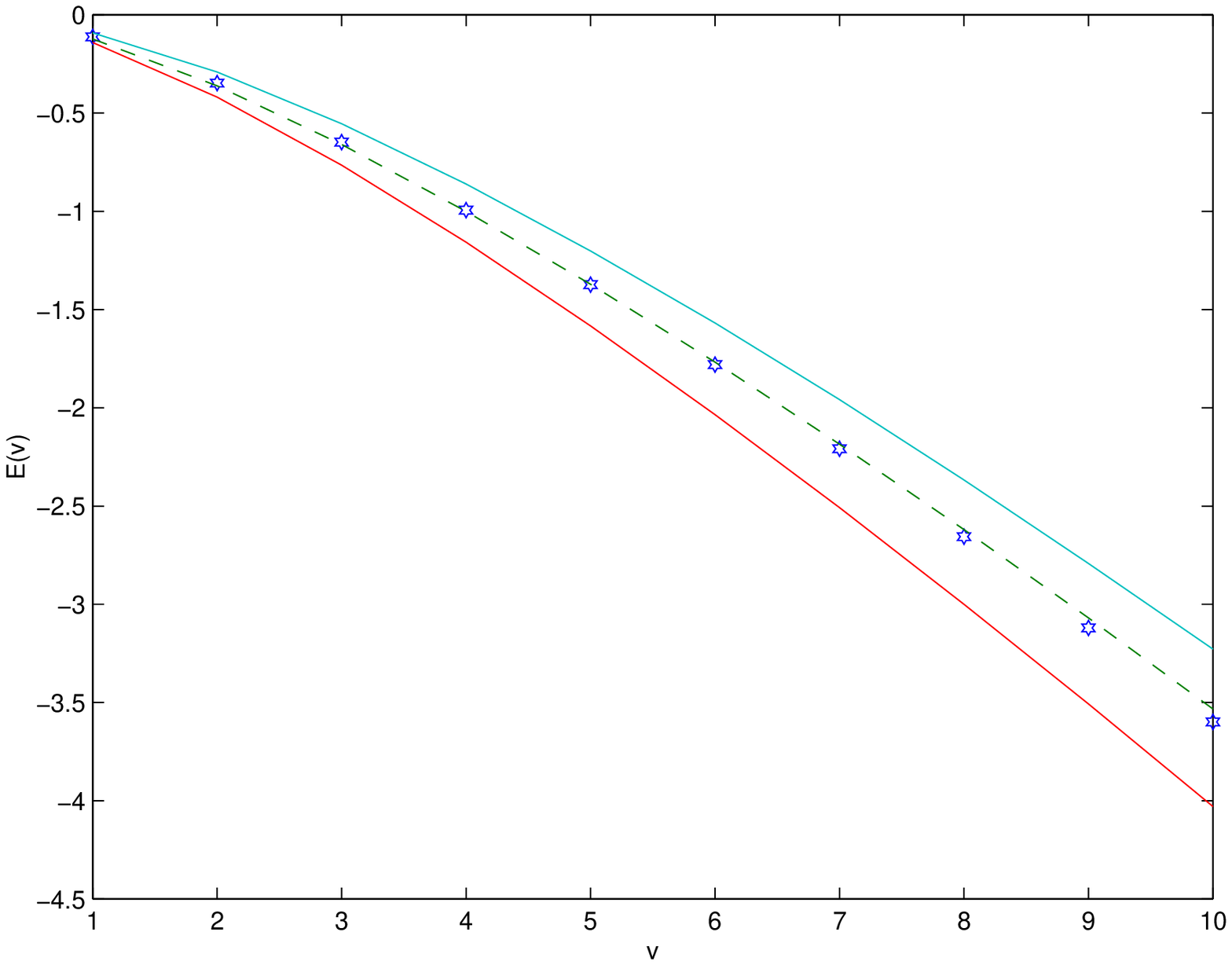,height=4in,width=3.5in,silent=}}}

\nl{\bf Figure 2.}~~Eigenvalue bounds(full-line)for the ground-state eigenvalue ${\cal E}(v)$ ($n = 1, \ell = 0$)of the Hamiltonian $H=-\Delta+vf(r)$ (in units $\hbar = 2m = 1$) for $b=1,$ together with accurate numerical data (dashed curve). The parametric equations (4.1) yield upper
 bounds when $P_{n\ell}=P_{n\ell}(1)$, lower bounds when $P_{n\ell}=P_{n\ell}(-1)$ and good approximation when $P_{n\ell} = \half(P_{n\ell}(-1)+P_{n\ell}(1)),$ shown as stars. 

\hfil\vfil
\end